\documentclass[onecolumn,10pt]{article}
\usepackage{amsmath}
\usepackage{amsfonts}
\usepackage{amssymb}
\usepackage{amsthm}
\usepackage{times}
\usepackage{cancel}
\usepackage[dvips]{graphics,color}
\newcommand{\hil}[1]{\mbox{$\mathcal{#1}$}}

\newcommand{\ket}[1]{| #1 \rangle}

\usepackage{tikz}

\title{\textbf{Von Neumann's `No Hidden Variables' Proof: A Re-Appraisal}\thanks{Dedicated to Peter Mittelstaedt, on the occasion of his 80'th birthday.}}
\author{Jeffrey Bub\\ \small \textit{Philosophy Department and Institute for Physical Science and Technology}\\  \small \textit{University of Maryland, College Park, MD 20742, USA}}

\date{}
\normalsize
\begin{document}

\maketitle

\begin{abstract}
Since the analysis by John Bell in 1965, the consensus in the literature
is that von Neumann's `no hidden variables' proof fails to exclude any significant
class of hidden variables. Bell raised the question whether it could be shown
that any hidden variable theory would have to be nonlocal, and in this sense
`like Bohm's theory.' His seminal result provides a positive answer to the
question. I argue that Bell's analysis misconstrues von Neumann's argument. What von Neumann proved was the impossibility of recovering the quantum probabilities from a hidden variable theory of dispersion free (deterministic) states in which the quantum observables are represented as the `beables' of the theory, to use Bell's term. That is, the quantum probabilities could not reflect the distribution of pre-measurement values of beables, but would have to be derived in some other way, e.g., as in Bohm's theory, where the probabilities are an artefact of  a dynamical process that is not in fact a measurement   of any beable of the system.
\end{abstract}

\bigskip

\section{Introduction}

John Bell began his analysis of of the hidden variable question in quantum mechanics with the following statement \cite[p.447]{Bellhv}:
\begin{quotation}
To know the quantum mechanical state of a system implies, in general, only statistical restrictions on the results of measurements. It seems interesting to ask if this statistical element be thought of as arising, as in classical statistical mechanics, because the states in question are averages over better defined states for which individually the results would be quite determined. These hypothetical `dispersion free' states would be specified not only by the quantum mechanical state vector but also by additional `hidden variables'--`hidden' because if states with prescribed values of these variables could actually be prepared, quantum mechanics would be observably inadequate. ... [The present paper] is addressed to those \ldots who believe that `the question concerning the existence of such hidden variables received an early and rather decisive answer in the form of von Neumann's proof on the mathematical impossibility of such variables in quantum theory.'
\end{quotation}

The reference is to a comment by Jauch and Piron \cite{JauchPiron1963}. Bell rejected von Neumann's argument as fatally flawed. In an interview for the magazine \emph{Omni} (May, 1988, p. 88), he went so far as to state:\footnote{Quoted by Mermin in \cite{Mermin1993}.}
\begin{quotation}
Yet the von Neumann proof, if you actually come to grips with it, falls apart in your hands! There is \emph{nothing} to it. It's not just flawed, its \emph{silly}! \ldots When you translate [his assumptions] into terms of physical disposition, they're nonsense. You may quote me on that: The proof of von Neumann is not merely false but \emph{foolish}!
\end{quotation}
My aim in this paper is to show that Bell's analysis misconstrues the nature of von Neumann's claim, and that von Neumann's argument actually establishes something important about hidden variables and quantum mechanics. What von Neumann proved was the impossibility of recovering the quantum probabilities from a hidden variable theory of dispersion free (deterministic) states  in which the quantum observables are represented as the `beables' of the theory, to use Bell's suggestive term.\footnote{Bell \cite[p.174]{BellBeables}: ``The beables of the theory are those elements which might correspond to elements of reality, to things which exist. Their existence does not depend on `observation.' Indeed observation and observers must be made out of beables.''}  That is, the quantum probabilities could not reflect the distribution of pre-measurement values of beables, but would have to be derived in some other way, e.g., as in Bohm's theory, where the probabilities are an artefact of  a dynamical process that is not in fact a measurement   of any beable of the system.

\section{The Proof}

Von Neumann's proof is set out in Chapter 4 of his book on quantum mechanics \cite{Neumann}. The analysis does not begin with a summation of the principles of quantum mechanics, but with the notion of a physical system as characterized by a set of `measurable quantities' and their functional relations. From the start, von Neumann is careful to distinguish simultaneously measurable quantities from quantities that are not simultaneously measurable \cite[p. 297]{Neumann}:
\begin{quotation}
Let us forget the whole of quantum mechanics but retain the following. Suppose a system $\mathbf{S}$ is given, which is characterized for the experimenter by the enumeration of all the effectively measurable quantities in it and their functional relations with one another. With each quantity we include the directions as to how it is to be measured---and how its value is to be read or calculated from the indicator positions on the measuring instruments. If $\hil{R}$ is a quantity and $f(x)$ any function, then the quantity $f(\hil{R})$ is defined as follows: To measure $f(\hil{R})$, we measure $\hil{R}$ and find the value $a$ (for $\hil{R}$). Then $f(\hil{R})$ has the value $f(a)$. As we see, all quantities $f(\hil{R})$ ($\hil{R}$ fixed, $f(x)$ an arbitrary function) are measured simultaneously with $\hil{R}$ \ldots But it should be realized that it is completely meaningless to try to form $f(\hil{R}, \hil{S})$ if $\hil{R}, \hil{S}$ are not simultaneously measurable: there is no way of giving the corresponding measuring arrangement.
\end{quotation}

Von Neumann alternates between the term `measurable quantity' and `physical quantity.' It is clear from the discussion that, in the context of the hidden variable question, he has in mind something like Bell's notion of `beable,' rather than `observable,' the standard terminology of quantum mechanics. Observables are usually identified with their representative Hermitian operators, but in the following it will be important to keep the two notions distinct. So I will use von Neumann's term `physical quantity'  and distinguish physical quantities, denoted by $\hil{A}, \hil{B}, \hil{S}_{x}$, etc., from Hermitian operators, denoted by $A, B, \sigma_{x}$, etc. To say that a system has a certain property is to associate a value, or a range of values, of a physical quantity with a system. 

The statistical properties of a statistical ensemble of copies of a system $\mathbf{S}$ are completely characterized by an expectation value function, defined for all  physical quantities of the system. Two definitions  \cite[p. 312]{Neumann}, labelled $\mathbf{\alpha'}, \mathbf{\beta'}$, define the notion of a dispersion free or deterministic ensemble, and a homogeneous or pure ensemble. An ensemble is \emph{dispersion free} or \emph{deterministic} if, for each quantity $\hil{R}$, $\mbox{Exp}(\hil{R}^{2}) = (\mbox{Exp}(\hil{R}))^{2}$, so the probabilities are all 0 or 1, and \emph{homogeneous} or \emph{pure} if $\mbox{Exp}(\hil{R})$ cannot be expressed as a convex combination of two different expectation value functions, neither of them equal to $\mbox{Exp}(\hil{R})$.

For von Neumann, the question of hidden variables underlying the quantum probabilities is the question of whether a quantum system can be regarded as characterized by certain parameters in addition to the quantum state, such that a specification of the values of these parameters in addition to the quantum state would `give the values of all physical quantities exactly and with certainty,' as von Neumann puts it in a preliminary discussion in Chapter 3 (p. 209). That is, the question is whether the quantum probabilities, for physical quantities construed as beables, can be derived by averaging over the distribution of dispersion free states associated with a given quantum state.

Von Neumann introduces two assumptions for physical quantities \cite[p. 311]{Neumann},  labelled $\mathbf{A'}, \mathbf{B'}$:
\begin{itemize}
\item[$\mathbf{A'}$:] If the quantity $\hil{R}$ is by nature non-negative, for example, if it is the square of another quantity $\hil{S}$, then $\mbox{Exp}(\hil{R}) \geq 0$
\item[$\mathbf{B'}$:] If $\hil{R}, \hil{S}, \ldots$ are arbitrary quantities and $a, b, \ldots$ real numbers, then $\mbox{Exp}(a\hil{R} + b\hil{S} + \ldots) = a\mbox{Exp}(\hil{R}) + b\mbox{Exp}(\hil{S}) + \ldots$
\end{itemize}

Now, von Neumann emphasizes \cite[p. 309]{Neumann} that if the quantities $\hil{R}, \hil{S}, \ldots$ are simultaneously measurable, then $a\hil{R} + b\hil{S} + \ldots$ is the ordinary sum, but if they are not simultaneously measurable, then `the sum is characterized \ldots only in an implicit way.' As an example \cite[pp. 309--310, footnote 164]{Neumann}, he refers to the kinetic energy of an electron in an atom, which is a sum of two quantities represented by noncommuting operators, a function of momentum, $\hil{P}$, and a function of position, $\hil{Q}$: 
\[
\hil{E} = \frac{\hil{P}^{2}}{2m} + V(\hil{Q})
\]
and he observes that we measure the energy by the measuring the frequency of the spectral lines in the radiation emitted by the electron, not by measuring the electron's position and momentum, computing the values of $\hil{P}^{2}/2m$ and $V(\hil{Q})$, and adding the results. 

Clearly, assumption $\mathbf{B'}$ is to be taken as a \emph{definition} of the physical quantity $a\hil{R} + b\hil{S} + \ldots$ when the quantities $\hil{R}, \hil{S}, \ldots$ are not simultaneously measurable, i.e., as defining a physical quantity, $\hil{X} \equiv a\hil{R} + b\hil{S} + \ldots $, whose expectation value is the sum $a\mbox{Exp}(\hil{R}) + b\mbox{Exp}(\hil{S}) + \ldots$ in all statistical ensembles (i.e., for all expectation value functions).

Two further assumptions \cite[pp. 313--314]{Neumann}, $\mathbf{I}, \mathbf{II}$, relate physical quantities to Hilbert space operators. It is assumed that each physical quantity of a quantum mechanical system is represented by a (hypermaximal)\footnote{Essentially, a hypermaximal operator is the most general Hermitian operator that has a spectral resolution.}  Hermitian operator in a Hilbert space, and that:
\begin{itemize}
\item[$\mathbf{I}$:] If the quantity $\hil{R}$ has the operator $R$, then the quantity $f(\hil{R})$ has the operator $f(R)$.
\item[$\mathbf{II}$:] If the quantities $\hil{R}, \hil{S}, \ldots$ have the operators $R, S, \ldots$, then the quantity $\hil{R} + \hil{S} + \dots$ has the operator $R+S+ \ldots$.
\end{itemize}

On the basis of the four assumptions $\mathbf{A'}$, $\mathbf{B'}$, $\mathbf{I}$, $\mathbf{II}$, and the two definitions $\mathbf{\alpha'}, \mathbf{\beta'}$, von Neumann proved that the expectation value function is uniquely defined by the trace function:
\begin{equation}
\mbox{Exp}(\hil{R}) = \mbox{Tr}(WR)
\end{equation}
where $W$ is a Hermitian operator independent of $R$ that characterizes the ensemble. It follows that there are no dispersion free ensembles, and that there are  homogeneous or pure ensembles, corresponding to the pure states of quantum mechanics. 

As is well-known, Gleason's theorem \cite{Gleason} later established the same result on the basis of weaker assumptions for Hilbert spaces of more than two dimensions: in effect, $\mathbf{II}$ is required to hold only for simultaneously measurable quantities. 

\section{Bell's Critique}

There is a quick argument that dispersion free ensembles are excluded if we identify the physical quantities of a hidden variable reconstruction of the quantum statistics with Hermitian operators, as $\mathbf{I}$ and $\mathbf{II}$ require, and assume that the possible values of a physical quantity are the eigenvalues of the representative operator.  As Bell pointed out in \cite{Bellhv}, for dispersion free states the expectation value of a physical quantity is equal to the eigenvalue in the dispersion free state, and it is clearly false that the eigenvalue of a sum of noncommuting operators is equal to the sum of the eigenvalues of the summed operators. For a spin-1/2 particle, for example, the eigenvalues of $\sigma_{x}$ and $\sigma_{y}$ are both $\pm 1$, while the eigenvalues of $\sigma_{x} + \sigma_{y}$ are $\pm \sqrt{2}$, so the relation cannot hold. 

Now, the same argument would apply to von Neumann's example of kinetic energy as a sum of a function of position and a function of momentum. As von Neumann asserts in the footnote mentioned in the previous section, the existence of  dispersion free or deterministic states is inconsistent with assumption $\mathbf{II}$  for quantities that are not simultaneously measurable. For example, if $\hil{R} = \hil{P}^{2}/2m$ and $\hil{S} = V(\hil{Q})$ are represented by the noncommuting operators $R = P^{2}/2m$ and $S = V(Q)$, then the  value of the quantity $\hil{E} = \hil{R} + \hil{S} = \hil{P}^{2}/2m +  V(\hil{Q})$ representing the kinetic energy of the system could not be the sum of the values of $\hil{R}$ and $\hil{S}$ in a dispersion free state, if $\hil{E}$ is represented by  the operator $R+S = P^{2}/2m + V(Q)$. So Bell's observation could not have escaped von Neumann. 

What the quick argument shows is that we cannot identify the physical quantities of a hidden variable theory with Hermitian operators, according to $\mathbf{I}$, $\mathbf{II}$, if we require the existence of dispersion free states. What the argument does \emph{not} show is what sorts of states are allowed for `physical quantities' in a generalized sense characterized by the conditions $\mathbf{A'}$, $\mathbf{B'}$, $\mathbf{I}$, $\mathbf{II}$, and this is clearly the more interesting question. Von Neumann's proof is  designed to answer this question, i.e., to derive  the full convex set of quantum probability distributions, and once this question is answered, the quick argument is redundant.

In \cite[p. 449]{Bellhv}, Bell characterizes von Neumann's proof as follows:
\begin{quotation}
Thus the formal proof of von Neumann does not justify his informal conclusion: `It is therefore not, as is often assumed, a question of reinterpretation of quantum mechanics---the present system of quantum mechanics would have to be objectively false, in order that another description of the elementary processes than the statistical one be possible.' It was not the objective measurable predictions of quantum mechanics which ruled out hidden variables. It was the arbitrary assumption of a particular (and impossible) relation between the results of incompatible measurements either of which might be made on a given occasion but only one of which can in fact be made.
\end{quotation}

The `arbitrary assumption of a particular (and impossible) relation' is the assumption that the expectation value of a sum of measurable quantities is the sum of the expectation values of the summed quantities, for quantities that are not simultaneously measurable, represented by noncommuting operators, as well as simultaneously measurable quantities represented by commuting operators, i.e., $\mathbf{B'}$ together with $\mathbf{II}$. The quotation suggests that von Neumann concluded that `the objective measurable predictions of quantum mechanics,' i.e., the quantum statistics, rule out hidden variables. This is misleading. In the comments immediately preceding this quotation, von Neumann writes \cite[pp. 324--325]{Neumann}:
\begin{quotation}
It should be noted that we need not go any further into the mechanism of the `hidden parameters,' since we now know that the established results of quantum mechanics can never be re-derived with their help. In fact, we have even ascertained that it is impossible that the same physical quantities exist with the same function connections (i.e., that $\mathbf{I}, \mathbf{II}$ hold) if other variables (i.e., `hidden parameters') should exist in addition to the wave function. 

Nor would it help if there existed other, as yet undiscovered, physical quantities, in addition to those represented by the operators in quantum mechanics, because the relations assumed by quantum mechanics (i.e., $\mathbf{I}, \mathbf{II}$) would have to fail already for the by now known quantities, those that we discussed above. It is therefore not, as is often assumed, \ldots
\end{quotation}

So the sense in which `the present system of quantum mechanics would have to be objectively false' if the quantum statistics could be derived from a distribution of dispersion free or deterministic states is that, in a hidden variable theory, \emph{the association of known physical quantities---like energy, position, momentum---with Hermitian operators in Hilbert space would have to fail.}

According to Bell, von Neumann proved only the impossibility of hidden deterministic states that assign values to a sum of physical quantities, $\hil{R} + \hil{S}$, that are the sums of the values assigned to the quantities $\hil{R}$ and $\hil{S}$, even when $\hil{R}$ and $\hil{S}$ cannot be measured simultaneously. As we saw, von Neumann regarded a sum of physical quantities that cannot be measured simultaneously as implicitly defined by the statistics, and he drew the conclusion that such an implicitly defined physical quantity  cannot be represented by the operator sum in a hidden variable theory.  

In \cite[p. 448]{Bellhv}, Bell constructed a toy hidden variable theory for spin-1/2 systems, i.e., for quantum systems represented on a 2-dimensional Hilbert space, in which eigenvalues are assigned to the spins in all
directions, given a quantum pure state $\ket{\psi}$ and a value of the hidden variable. On the face of it, this would seem to be a counterexample to von Neumann's theorem.

Now in Bell's example, there is an implicitly defined physical quantity in von Neumann's sense, $\hil{S} = \hil{S}_{x}+ \hil{S}_{y}$, whose actual value, in a dispersion free or deterministic state, is  the sum of the actual values of $\hil{S}_{x}$ and $\hil{S}_{y}$. So, trivially, $\mbox{Exp}(\hil{S}) = \mbox{Exp}(\hil{S}_{x}) + \mbox{Exp}(\hil{S}_{y})$ for all statistical ensembles, i.e., for the expectation values calculated by averaging over the dispersion free or deterministic states (as opposed to expectation values calculated via the trace formula for the representative operators). If, with von Neumann, we `forget the whole of quantum mechanics,' an argument is required for taking the sum of the Hermitian operators $\sigma_{x}, \sigma_{y}$, with eigenvalues $\pm\sqrt{2}$, as representing the physical quantity $\hil{S}$. Similarly, some explanation is necessary to account for the applicability of the trace formula to the Hermitian operator $\sigma_{x} + \sigma_{y}$ in generating the quantum statistics of the physical quantity $\hil{S}$, i.e., the measurement statistics of the implicitly defined sum of the physical quantities $\hil{S}_{x}$ and $\hil{S}_{y}$. This is precisely what Bohm does in his hidden variable theory \cite{Bohm2}, on the basis of a disturbance theory of measurement that  generates the quantum statistics.

For example, the momentum of a Bohmian particle is the rate of change of position, but the expectation value of momentum in a quantum ensemble is not derived by averaging over the particle momenta. Rather, the expectation value is derived  via a theory of measurement, which yields the trace formula involving the momentum \emph{operator} if we assume, as a contingent fact, that the probability distribution of hidden variables---particle positions---has reached equilibrium. Bohm \cite[p 387]{Bohm2} gives an example of a free particle in a box of length $L$ with perfectly reflecting walls. Because the wave function is real, the particle is at rest. The kinetic energy of the particle is $E = P^{2}/2m = (nh/L)^{2}/2m$. Bohm asks: how can a particle with high energy be at rest in the empty space of the box? The solution to the puzzle is that a measurement of the particle's momentum changes the wave function, which plays the role of a guiding field for the particle's motion, in such a way that the \emph{measured} momentum values will be $\pm nh/L$ with equal probability. Bohm comments \cite[pp. 386--387]{Bohm2}:
\begin{quotation}
This means that the measurement of an `observable' is not really a measurement of any physical property belonging to the observed system alone. Instead, the value of an `observable' measures only an incompletely predictable and controllable potentiality belonging just as much to the measuring apparatus as to the observed system itself.
\ldots
We conclude then that this measurement of the momentum `observable' leads to the same result as is predicted in the usual interpretation. However, the actual particle momentum existing before the measurement took place is quite different from the numerical value obtained for the momentum `observable,' which, in the usual interpretation, is called the 'momentum.'
\end{quotation}

In Bohm's hidden variable theory, physical quantities are introduced as classical dynamical quantities, functions of positions and momenta, i.e., as beables, and the quantum statistics defined by the trace rule for quantum observables is an artefact  of  a dynamical process that is not in fact a measurement of any physical quantity of the system. For von Neumann, the physical quantities of quantum systems and their functional relations are abstracted from the measurement statistics and appropriately represented by Hermitian operators, and the quantum theory, including the quantum theory of measurement, is a theory about these physical quantities. 

The pre-dynamic structure of quantum mechanics, given by the Hilbert space representation of physical quantities as Hermitian operators and the trace formula for expectation values, excludes hidden variables in a similar sense to which the pre-dynamic structure of special relativity, given by the non-Euclidean geometry of Minkowski space-time, excludes hidden forces as responsible for Lorentz contraction. Lorentz aimed to maintain Euclidean geometry and Newtonian kinematics and explain the anomalous behavior of light in terms of a dynamical theory about how rods contract as they move through the ether. His dynamical theory plays a similar role, in showing how Euclidean geometry can be preserved, as Bohm's dynamical theory of measurement in explaining how the quantum statistics can be generated in a classical or Boolean theory of probability.

\section{Conclusion}

Von Neumann's proof establishes that if the physical quantities of quantum mechanics, like energy, position, momentum, etc., are characterized by the conditions $\mathbf{A'}$, $\mathbf{B'}$, $\mathbf{I}$, $\mathbf{II}$, then dispersion free states are excluded and the quantum pure states  are the appropriate extremal states for quantum probability distributions. So in a hidden variable theory in which dispersion free (deterministic) states are the extremal states for quantum probability distributions, the quantum probabilities could not reflect the distribution of pre-measurement values of beables, but would have to be derived in some other way, , e.g., as in Bohm's theory, where the probabilities are an artefact of  a dynamical process that is not in fact a measurement   of any beable of the system. What von Neumann's proof excludes, then, is the class of hidden variable theories in which (i) dispersion free (deterministic) states are the extremal states, and (ii) the beables of the hidden variable theory correspond to the physical quantities represented by the Hermitian operators of quantum mechanics.

\section*{Acknowledgements}

Research supported by  the University of Maryland Institute for Physical Science and Technology. I  thank  Michael Cifone for insightful discussions on Bohm's theory.

\bibliographystyle{plain}
\bibliography{vn}

\end{document}